\documentclass{ws-procs9x6}

\begin{document}

\title{BOSONIZATION OF THE SCHWINGER MODEL\\
BY NONCOMMUTATIVE CHIRAL BOSONS}

\author{J. BEN GELOUN,$^{\dagger}$  J. GOVAERTS$^{\dagger,\ddagger}$ and
M.~N. HOUNKONNOU$^{\dagger}$}

\address{$^{\dagger}$International Chair in Mathematical Physics 
and Applications (ICMPA),\\
072 B.P. 50, Cotonou, Republic of Benin\\
E-mail: jobengeloun@yahoo.fr, norbert\_hounkonnou@cipma.net}

\address{$^\ddagger$Center for Particle Physics and Phenomenology (CP3),\\
Institute of Nuclear Physics, Catholic University of Louvain,\\
2, Chemin du Cyclotron, B-1348 Louvain-la-Neuve, Belgium\\
E-mail: Jan.Govaerts@fynu.ucl.ac.be}

\begin{abstract}
Bosonization of the Schwinger model with noncommutative chiral bosons is 
considered on a spacetime of cylinder topology. Using point splitting 
regularization, manifest gauge invariance is maintained throughout.
Physical consequences are discussed.
\end{abstract}

\vspace{50pt}

\begin{center}
To appear in the Proceedings of the Fourth International Workshop on Contemporary
Problems in Mathematical Physics,\\
November 5$^{\rm th}$-11$^{\rm th}$, 2005, Cotonou (Republic of Benin),\\
eds. J. Govaerts, M. N. Hounkonnou and A. Z. Msezane\\
(World Scientific, Singapore, 2006).
\end{center}

\vspace{50pt}

\begin{flushright}
CP3-06-09\\
ICMPA-MPA/2006/26
\end{flushright}

\bodymatter

\clearpage

\section{Introduction}
\label{Sec1}

Bosonization\cite{col} proves to be a successful method for
quantization in noncommutative field theory.\cite{da}$^{-}$\cite{do}
On the other hand, it is well known that the ordinary massless Schwinger model
is exactly soluble through the bosonization of the massless fermion,
in particular within the physical projector method which avoids having to
perform any gauge fixing procedure.\cite{ag} In this contribution, 
we investigate a new quantized version of the Schwinger model via 
a noncommutative chiral bosonization, taking the spacelike dimension
to be compactified into a circle, ${\mathbb{R}} \to S^{1}$.
The quantization to be considered presently is realized in 
a noncommutative field space instead of a noncommutative spacetime.
The Hamiltonian analysis is significantly simplified without loss of 
a meaningful physical interpretation. The quantized system thereby
obtained generalizes the nonperturbative quantization
of the model\cite{ag} as an asymptotic $\theta$-quantization, and
the quantization rules found by Das {\it et al.}\cite{da}
are extended to a nontrivial spacetime topology.
Furthermore, we provide a gauge invariant quantized model by
the point splitting regularization method using the Wilson line phase factor.
The Hamiltonian operator is diagonalized
leading to a successful nonperturbative quantization.
Such an abelian gauge theory based on the circle $S^{1}$
as the space topology is new to the best of our knowledge.

Section~\ref{Sec2} presents our notations and briefly describes the classical 
Schwinger model and the relevant quantities of the constrained dynamics.
In Sec.~\ref{Sec3}, we study chiral bosons in the noncommutative field 
space and develop the nonperturbative quantization of the model based 
on the bosonization of the fermionic degrees of freedom. The gauge 
invariant regularization is given in Sec.~\ref{Sec4}. Finally, some
conclusions are presented in Sec.~\ref{Sec5}.

\vspace{10pt}

\section {The Classical Schwinger Model}
\label{Sec2}

Henceforth, the topology of the 1+1 dimensional spacetime is
that of the cylinder $\mathbb{R} \times S^{1}$, where $\mathbb{R}$ stands 
for the timelike coordinate and the torus $S^{1}$ of radius $R$ and 
circumference $L=2\pi R$ for the spacelike one. The Minkowski metric 
is $\eta_{\mu\nu}= {\rm diag} (+,-)$, $\mu,\nu = 0,1$. Units such
that $\hbar=1=c$ are used throughout. The antisymmetric tensor 
$\epsilon^{\mu\nu}$ is such that $\epsilon^{01}=1=-\epsilon^{10}$.
In the chiral representation, the Clifford-Dirac algebra
$\left\{ \gamma^{\mu}, \gamma^{\nu} \right\}= 2\eta^{\mu\nu}$
is given by $\gamma^{0}=\sigma^{1}$, $\gamma^{1}= i\sigma^{2}$ and
$\gamma_{5}=\gamma^{0}\gamma^{1}=-\sigma ^{3}$, where 
$\sigma ^{i}$ $(i=1,2,3)$ are the usual Pauli matrices.

\clearpage4

The field degrees of freedom of the model are the real $U(1)$ gauge 
vector field $A_{\mu}(x^{\mu})$ and a single massless Dirac spinor 
$\psi(x^{\mu})$ represented by Grassmann odd variables,
describing the fermionic particles. The Dirac spinor decomposes 
into two complex Weyl spinor representations of opposite chiralities,
$\psi(t,x)= \psi_{+}(t+x) + \psi_{-}(t-x)$, such that 
$\gamma_{5}\psi_{\pm}= \mp\psi_{\pm}$. Furthermore, the following 
choice of periodic and twisted (or anti-periodic)
boundary conditions on the circle is assumed, respectively,
\begin{eqnarray}\label{bound}
A_{\mu}(t, x+L)=A_{\mu}(t,x),\qquad
\psi_{\pm}(t, x+L)= - e ^{2i \pi\alpha_{\pm}} \psi_{\pm}(t,x),
\end{eqnarray}
$\alpha_{\pm}$ being arbitrary real parameters
defined modulo any integer such that
$\alpha_{+}=\alpha_{-}=\alpha\ ({\rm mod}\, \mathbb{Z})$
in order to ensure parity invariance.

The model is described by the Lagrangian density given
in a Lorentz and gauge invariant form as (Einstein's summation convention 
is implicit)
\begin{eqnarray}\label{lagr}
{\mathcal L}= -\frac{1}{4}F_{\mu\nu}F^{\mu\nu} + 
\frac{1}{2}i \bar\psi\gamma^{\mu}
( \partial_{\mu} + ie A_{\mu})\psi
-\frac{1}{2}i(\partial_{\mu} -ie A_{\mu})\bar\psi\gamma^{\mu}\psi,
\end{eqnarray}
with $F_{\mu\nu}= \partial_{\mu}A_{\nu}-\partial_{\nu}A_{\mu}$,
the gauge field strength corresponding, in $1+1$ dimensions, 
to a single field $F_{01}$ which is the pseudo-scalar
electric field $E$; $e$ stands for the gauge coupling constant which,
up to a sign, is the charge of the fermionic particle 
(electron or positron).

Gauge invariance of the system implies a constrained dynamics\cite{ag} whose
Dirac Hamiltonian treatment leads to a first-class constraint $\Phi$ 
(Gauss' law) and the first-class Hamiltonian density ${\mathcal H}$
given as follows,
\begin{eqnarray}\label{firstcc}
\Phi&=& \partial_{1}\pi_{1} + e\psi^{\dag}\psi,\;\; \pi_{1}=-E, \\
\label{hamilton}
{\mathcal H}&=& \frac{1}{2} \pi_{1}^{2}
- \frac{1}{2}i \psi^{\dag}\gamma_{5}( \partial_{1} - ie A^{1})\psi
+ \frac{1}{2}i ( \partial_{1} + ie A^{1})\psi^{\dag}\gamma_{5}\psi,
\end{eqnarray}
where $\pi_{1}$ is the momentum conjugate to $A^{1}$. In this system,
$A^{0}$ turns out to be a Lagrange multiplier for $\Phi$.
The fundamental graded Dirac-Poisson algebra, taken at equal time, 
reads ($a, b= +,- $),
\begin{eqnarray}\label{commut}
\{ A^{1}(t,x),\pi_{1}(t,y) \}=\delta(x-y),\quad
\{ \psi_{a}(t,x), \psi^{\dag}_{b}(t,y) \}= -i\delta_{ab}\delta(x-y).
\end{eqnarray}

\section{Nonperturbative Deformed Quantization}
\label{Sec3}

Bosonization of the fermionic sector of the Schwinger model involves
the quantum deformed algebra of chiral bosons, $\phi_+(t, x)$
and $\phi_-(t,x)$. In the Schr\"odinger picture at time $t=0$,
we have the Fourier mode decomposition of the chiral fields $\phi_\pm$,

\clearpage

\begin{eqnarray}
\phi_{\pm}(x)&=&
 \frac{1}{\sqrt{4\pi}}
   \left\{
     q_{\pm} + \frac{2\pi}{L}p_{\pm}(\pm x)
    \right. \cr
&&\left.
+ \sum_{n=1}^{+\infty} \frac{1}{\sqrt{n}}
     \left[ {\bf a}_{\pm, n} e^{-\frac{2i\pi}{L} n(\pm x) } +
      {\bf a}_{\pm, n}^{\dag}e^{\frac{2i\pi}{L} n(\pm x) }
      \right]
 \right\}.
\end{eqnarray}
Representing a noncommutative field space, the deformed Dirac algebra 
of the chiral bosons proposed by Das {\it et al.}\cite{da}
induces the following quantum mode algebras corresponding to 
the chiral fields $\phi_\pm$,
\begin{eqnarray}\label{deform}
\left[ q_{a}, p_{b} \right] &=& i\Delta_{ab},\qquad
\left[ {\bf a}_{a, n}, {\bf a}_{b, m}^{\dag} \right]=
\delta_{ab}\delta_{n,m}, \cr
\left[ {\bf a}_{a, n}, {\bf a}_{b, m} \right]&=& 
-\epsilon_{ab}\theta\delta_{n,m}=
-\left[ {\bf a}_{a, n}^{\dag}, {\bf a}_{b, m}^{\dag} \right],
\end{eqnarray}
with $a,b = +,-$, $\theta$ a real parameter, and 
$\Delta_{ab}= \epsilon_{ab}\theta + \delta_{ab}$.
Furthermore, we assume the hermiticity conditions
for the zero modes $q_{a}=q^{\dag}_{a}$ and $p_{a}=p^{\dag}_{a}$.

In terms of these bosonic modes, the fermionic operator 
$\psi_{a}$ is written as $\psi_{a}(z)= F(z)V_{a}(z)$ with $V_{a}$
the vertex operator defined as
$$V_{a}=K(a):e^{ai \beta (\varphi_{a} -a \theta \varphi_{-a})}:,$$
where $F(z)$ is a holomorphic function, $\varphi_{a}=\sqrt{4\pi}\phi_{a}$ and
$K(a)$ is fixed to $K(a)=e^{\frac{i\pi}{2}\rho_{a}p_{-a}}$.
The constant factors $\beta$ and  $\rho_{a}$ are determined by 
the bosonization process. The vertex operators $V_{a}$ and $V_{a}^{\dag}$ 
are written as Laurent series as
\begin{eqnarray}\label{vertex1}
V_{\pm}(z)= \sum_{n=-\infty}^{\infty}\psi_{\pm, n} 
z^{ - (n \mp \frac{1}{2} \mp \alpha_{\pm}) - \frac{1}{2} }, \\
V_{\pm}^{\dag}(z)= \sum_{n=-\infty}^{\infty}\psi_{\pm, n}^{\dag} 
z^{ (n \mp \frac{1}{2} \mp \alpha_{\pm}) - \frac{1}{2} }.
\label{vertex2}
\end{eqnarray}
{}From (\ref{vertex1}) and (\ref{vertex2}), 
we deduce, by a complex contour integral 
around the origin $z=0$, the expressions for the mode operators 
$\psi_{\pm, n}$ and $\psi_{\pm, n}^{\dag}$ as
\begin{eqnarray}
\psi_{\pm, n}= \oint_{C} \frac{dz}{2i\pi}
z^{ (n \mp \frac{1}{2} \mp \alpha_{\pm}) - \frac{1}{2} }V_{\pm}(z), \quad
\psi_{\pm, n}^{\dag}= \oint_{C} \frac{dz}{2i\pi}
 z^{ - (n \mp \frac{1}{2} \mp \alpha_{\pm}) - \frac{1}{2} }V_{\pm}^{\dag}(z).
\nonumber\end{eqnarray}
By well established conformal field theory techniques,\cite{gi}
the fermionic mode anticommutators are evaluated. One has,
\begin{eqnarray}\label{psipsidag}
\left[ \psi_{a, n}, \psi_{b, m}^{\dag}\right]_{+} &=&
\oint_{C} \frac{dz}{2i\pi}  \oint_{C} \frac{d\omega}{2i\pi}
z^{ N - \frac{1}{2} }
\omega ^{ -M - \frac{1}{2} } 
\left[ V_{a}(z), V_{b}^{\dag}(\omega) \right]_{+} \cr
&=&\oint_{C} \frac{dz}{2i\pi}z^{N-\frac{1}{2}}
\oint_{z}\frac{d\omega}{2i\pi}
\omega^{-M-\frac{1}{2}} \times R(V_{a}(z)V_{b}^{\dag}(\omega)),
\end{eqnarray}
with $a,b= +,-$, $N=n-a(\frac{1}{2}+\alpha_a)$ and
$M=m-b(\frac{1}{2}+\alpha_b)$.
%$N = n \mp \frac{1}{2} \mp \alpha_{\pm}$ and
%$M= m \mp \frac{1}{2} \mp \alpha_{\pm}$.
We have used the Wick theorem in (\ref{psipsidag}). Integration
over $\omega$ stands for an integration depending on $z$, while
the last integral over $z$ is taken around the origin, $z =0$.
Using $:e^{A}::e^{B}:= e^{<AB>}: e^{A}e^{B} :$
where $A, B$ are some operators whose commutator is a $c$-number, 
we determine the radial ordered product
$R(V_{a}(z)V_{b}^{\dag}(\omega))$ as follows,
\begin{eqnarray}\label{rordered}
&&R(V_{a}(z)V_{b}^{\dag}(\omega)) \cr
&=&\left[\Theta (\mid z\mid-\mid\omega\mid)
  \left(\frac{1}{ (z-\omega)^{\delta_{ab}} }z^{-\theta\epsilon_{ab}}
                \right)^{\beta^{2}(1+ \theta^{2})}
  e^{a\beta\rho_{a}\frac{i\pi}{2}(1+ \theta^{2})\delta_{a(-b)} }\right.
    \cr
& &\left. - \Theta(\mid\omega\mid-\mid z\mid)
    \left(\frac{1}{(\omega-z)^{\delta_{ab}}}\omega^{\theta\epsilon_{ab}}
               \right)^{\beta^{2}(1+ \theta^{2})}
  e^{-a\beta\rho_{-a}\frac{i\pi}{2}(1+\theta^{2})\delta_{a(-b)} }\right]
\cr
&&\times
:e^{i\frac{\pi}{2}(\rho_{a}p_{-a}-\rho_{b}p_{-b})
+ i\beta(a(\varphi_{a}- a\theta\varphi_{-a})(z)-b(\varphi_{b}-
b\theta\varphi_{-b})(\omega))}:,
\end{eqnarray}
where $\Theta(\cdot)$ denotes the Heaviside step function.
The different parameters are fixed by
\begin{eqnarray}\label{parameters}
\beta=\pm\lambda(1+ \theta^{2})^{-\frac{1}{2}},\quad
\rho_{a}=\rho_{-a}=\rho(1+ \theta^{2})^{-\frac{1}{2}},\quad
{\mbox {with}}\;\lambda^{2} =1= \rho^{2}.
\end{eqnarray}
The induced singular part of the short distance operator ,
as $z\to \omega$, reads
\begin{eqnarray}\label{ope}
R(V_{a}(z)V_{a}^{\dag}(\omega))= \frac{1}{(z-\omega)} + \cdots
\end{eqnarray}
so that, by the residue theorem and (\ref{ope}), we get
\begin{eqnarray}\label{psipsidag2}
\left[ \psi_{a, n}, \psi_{b, n}^{\dag}\right]_{+} = \delta_{ab}\delta_{nm},
\end{eqnarray}
which reproduces the well known anticommutator of the corresponding 
fermionic operators. In the same manner, given (\ref{parameters}), 
evaluating the other radial ordered products 
$R(V_{a}^{\dag}(z)V_{b}(\omega))$, $R(V_{a}(z)V_{b}(\omega))$ and
$R(V_{a}^{\dag}(z)V_{b}^{\dag}(\omega))$ and then integrating the corresponding
fermionic algebra, the remaining fermionic mode anticommutators are recovered
as follows,
\begin{eqnarray}
\left[ \psi_{a, n}^{\dag}, \psi_{b, n}\right]_{+} = \delta_{ab}\delta_{nm},
\quad
\left[ \psi_{a, n}, \psi_{b, n}\right]_{+}=0,\quad
\left[ \psi_{a, n}^{\dag}, \psi_{b, n}^{\dag}\right]_{+} = 0.
\end{eqnarray}
Indeed, as $z \to \omega $, the only singularity such as the one in (\ref{ope})
with a nonvanishing residue occurs for $R(V_{a}^{\dag}(z)V_{b}(\omega))$.
We complete this study by giving the bosonized fermionic field 
(the $\theta$-Mattis--Mandelstam formula),

\clearpage

\begin{eqnarray}\label{bosonized}
\psi_{\pm}(x)&=&
 \frac{1}{\sqrt{L}}e^{\pm\frac{i\pi}{L}x}
e^{\frac{\pm i \lambda }{\sqrt{1+\theta^2 }} (q_{\pm}\mp \theta q_{\mp})}
 e^{\frac{i\pi}{2}\frac{\rho p_{\mp}}{\sqrt{1+\theta^2 }}}
 e^{\frac{2i\pi \lambda }{L\sqrt{1+\theta^2 }} (p_{\pm}\pm \theta p_{\mp})x} \cr
& & \times
\prod_{n=1}^{\infty}e^{\frac{\pm i \lambda }{\sqrt{n}\sqrt{1+\theta^2 }}
 ({{\bf a}}_{\pm,n}^{\dag}e^{\pm\frac{2i\pi}{L}n x}
 \mp \theta {{\bf a}}_{\mp,n}^{\dag}e^{\mp\frac{2i\pi}{L}n x})}   \cr
& & \times
\prod_{n=1}^{\infty}e^{\frac{\pm i \lambda }{\sqrt{n}\sqrt{1+\theta^2 }}
  ({{\bf a}}_{\pm,n}e^{\mp\frac{2i\pi}{L}n x}
  \mp \theta {{\bf a}}_{\mp,n}e^{\pm\frac{2i\pi}{L}n x})}.
 \nonumber
 \end{eqnarray}
The holonomy boundary condition (h.b.c.) of the chiral bosons
should also be related to $\alpha_{\pm}$, namely
the h.b.c. of the fermionic operator $\psi_{\pm}$,
through the deformed Heisenberg algebra on the circle.
This relation has still to be understood (this work is in progress).
Finally, the bosonic fields $\phi_{\pm}$ and their quantum states
also provide a representation of the fermionic algebra irrespective 
of the choices $\rho= \pm 1$ and $\lambda= \pm 1$.

\section{Gauge Invariant Regularization}
\label{Sec4}

The first-class quantities (\ref{firstcc}) and
(\ref{hamilton}) are composite operators which require
a choice of operator ordering which should remain anomaly free and 
gauge invariant. Since gauge invariance must be preserved at all steps, 
a gauge invariant point splitting regularization of short distance 
singularities is a relevant choice.\cite{ag} For instance, instead of 
the products $\psi_{\pm}^{\dag}(x)\psi_{\pm}(x)$,
one considers the gauge invariant quantities
$$\lim_{y \to x}\psi_{\pm}^{\dag}(y)e^{ie \int_{x}^{y} du  
A^{1}(u)}\psi_{\pm}(x)$$
and subtracts the divergent terms. Through this procedure,
the fermionic currents in the bosonized representation are given by
\begin{eqnarray}
\psi_{\pm}^{\dag}\psi_{\pm}= -\frac{\lambda}{2\pi\sqrt{1+\theta^{2}}}
\left[ \partial_{1}( \varphi_{\pm} \mp \theta \varphi_{\mp})
\mp \lambda e \sqrt{1+\theta^{2}}  A^{1} \right],
\end{eqnarray}
so that the vector and axial gauge invariant currents are expressed as, 
respectively,
\begin{eqnarray}\label{current}
\psi^{\dag}\psi &=&
-\frac{\lambda}{2\pi\sqrt{1+\theta^{2}}}
\left[\partial_{1}( \varphi_{+}+\varphi_{-})
+\theta\partial_{1}(\varphi_{+}-\varphi_{-})\right],\cr
\psi^{\dag}\gamma_{5}\psi &=&
-\frac{e}{\pi}\left[ A^{1} -\frac{1}{2\lambda e \sqrt{1+\theta^{2}}}
\left[ \partial_{1} (\varphi_{+}-\varphi_{-}) - 
\theta\partial_{1}(\varphi_{+} + \varphi_{-}) \right]\right].
\end{eqnarray}
Denoting the regularized current components by
$J^{\pm}= \psi_{\pm}^{\dag}\psi_{\pm}$,
the two coupled $U(1)$ Kac-Moody algebras\cite{da}
$\left[ J^{a}, J^{b} \right] = (a\delta_{ab} + 
\epsilon_{ab}\theta)\frac{i}{2} \partial_{x}\delta(x-y)$
for the unmixed and mixed commutators
are still valid. The gauge contribution is indeed null.
Furthermore, we have the regularized vector and axial charges given by
\begin{eqnarray}
Q &=& -\frac{\lambda}{\sqrt{1+\theta^{2}}}
\left[ p_{+} - p_{-} + \theta (p_{+} + p_{-}) \right], \cr
Q_{5}&=&
-\frac{e}{\pi}\int_{0}^{L}dx A^{1} + \frac{\lambda}{\sqrt{1+\theta^{2}}}
 \left[ p_{+} + p_{-} - \theta (p_{+} - p_{-}) \right].
\end{eqnarray}
One notices that the quantum axial anomaly of the axial charge,
due to the zero mode of the gauge potential $A^{1}$, remains as in
the commutative case. The regularized first-class constraint reads
\begin{eqnarray}
\Phi= \partial_{1}\pi_{1}-
\frac{e\lambda}{2\pi\sqrt{1+\theta^{2}}}
\left[\partial_{1}(\varphi_{+}+\varphi_{-}) +
\theta\partial_{1}( \varphi_{+} - \varphi_{-}) \right].
\end{eqnarray}
The bilinears involving the covariant derivatives, namely,
$\psi_{\pm}^{\dag}D_{1}\psi_{\pm}$ and $D_{1}\psi_{\pm}^{\dag}\psi_{\pm}$, are
also regularized by the point splitting method.
Setting $D_{1}\psi=(\partial_{1}-ieA^{1})\psi$ and
$D_{1}\psi^{\dag}=(\partial_{1}+ieA^{1})\psi^{\dag}$,
we obtain the fermionic contributions to the first-class Hamiltonian
\begin{eqnarray}
H_{\pm} &=&  \frac{i}{2}\psi_{\pm}^{\dag}D_{1}\psi_{\pm}
 -\frac{i}{2}D_{1}\psi_{\pm}^{\dag}\psi_{\pm} \cr
&=& \pm \frac{1}{4\pi(1+\theta^{2})} (\partial_{1}
( \varphi_{\pm}\mp\theta\varphi_{\mp})
\mp \lambda e\sqrt{1+\theta^{2}} A^{1} )^{2} \mp \frac{\pi}{12 L^{2}},
\end{eqnarray}
showing that the deformation has not modified the Casimir energy
${\pi}/(12 L^{2})$.

The first-class gauge invariant Hamiltonian density reads
\begin{eqnarray}\label{fhamregular}
{\mathcal H}&=& \frac{1}{2}\pi_{1}^{2}+ \frac{1}{4\pi(1+\theta^{2})}
\left[   \left( \partial_{1}( \varphi_{+}- \theta\varphi_{-})
                - \lambda e \sqrt{1+\theta^{2}} A^{1}
         \right)^{2}
 \right.\cr
& & + \left.
         \left( \partial_{1} ( \varphi_{-} + \theta\varphi_{+} )
        + \lambda e\sqrt{1+\theta^{2}} A^{1}
         \right)^{2}
    \right].
\end{eqnarray}
Introducing the change of variables
\begin{equation}
\begin{array}{rcl}
\Upsilon & = &  \frac{-1}{\mu}\pi_{1}, \\
 & & \\
\pi_{\Upsilon} & = & \mu 
\left\{\left(A^{1} -\frac{1}{2e\lambda\sqrt{1+\theta^{2}}}
\partial_{1}\left[ (\varphi_{+} - \varphi_{-})
                   - \theta(\varphi_{+} + \varphi_{-}\right)\right]\right\},
\end{array}
\end{equation}
with $\mu={\mid e \mid}/{\sqrt{\pi}}$, the first-class Hamiltonian 
density (\ref{fhamregular}) can be expressed
using the new degrees of freedom $( \Upsilon, \pi_{\Upsilon}, \varphi_{a},
 \partial_{1}\varphi_{a}$) as
\begin{eqnarray}
{\mathcal H}=  \frac{1}{2}\pi_{\Upsilon}^{2} + 
\frac{1}{2}(\partial_{1}\Upsilon)^{2}
+ \frac{1}{2}\mu^{2}\Upsilon^{2} + 
\frac{1}{2}\left(\frac{\Phi}{\mu}\right)^{2}+
\left(\frac{\Phi}{\mu}\right)\partial_{1}\Upsilon.
\end{eqnarray}
Finally, given $\Delta_{ab}= \epsilon_{ab}\theta + \delta_{ab}$, 
we have the bosonic algebra
\begin{equation}
\begin{array}{rl}
& \left[ \Upsilon(x), \pi_{\Upsilon}(y)\right] = i\delta(x-y),\quad
\left[\varphi_{a}, \partial_{1}\varphi_{b} \right] =  2ib\pi\Delta_{ab}
\delta(x-y),\\
 & \\
& \left[ \varphi_{a}, \pi_{\Upsilon}(y) \right] = 
- \frac{i\pi\mu}{e\lambda}\sqrt{1+\theta^{2}}\delta(x-y).
\end{array}
\end{equation}

\section{Concluding Remarks}
\label{Sec5}

In this work, we have generalized to the cylinder spacetime topology
${\mathbb R} \times S^{1}$ and to the massless
Schwinger model the noncommutative bosonization induced by a deformed 
Dirac algebra of noncommutative chiral bosons. The usual quantum theory 
is recovered as $\theta \to 0$. This quantum model, regularized by 
the gauge invariant point splitting method, induces a gauge invariant 
dynamics equivalent to that of an electric field of mass 
$\mu={\mid e\mid}/{\sqrt{\pi}}$, like in the ordinary commutative case. 
Furthermore, due to the choice of the chiral boson algebra, a charge 
rescaling $e \to e\sqrt{1+ \theta^{2}}$ appears in the expressions.
This fact has still to be related to the equivalence found by 
Das {\it et al.}\cite{da} which states that noncommutative
chiral bosons are equivalent to free fermionic fields moving
with a speed equal to $c'=c\sqrt{1+ \theta^{2}}$, $c$ being the speed of 
light in vacuum. We expect that, in a forthcoming investigation,
a stronger deformation involving the nonzero mode algebra of the 
chiral fields would lead to a noncommutative spinor field theory.

\section*{Acknowledgments}

J.~B.~G. is grateful to the Abdus Salam International Centre for Theoretical
Physics (ICTP, Trieste, Italy) for a Ph.D. fellowship under the
grant \mbox{Prj-15}. J. G. acknowledges the Abdus Salam ICTP
Visiting Scholar Programme in support of a Visiting Professorship
at the International Chair in Mathematical Physics and Applications (ICMPA).
The work of J. G. is partially supported by the 
Belgian Federal Office for Scientific, Technical and Cultural Affairs
through the Interuniversity Attraction Pole (IAP) P5/27.

\end{document}